\documentclass[prb,twocolumn,showpacs,showkeys]{revtex4}
\usepackage{dcolumn}
\begin{document}
\newcommand{\be}{\begin{equation}}
\newcommand{\ee}{\end{equation}}
\newcommand{\ba}{\begin{eqnarray}}
\newcommand{\ea}{\end{eqnarray}}
\newcommand{\vk}{{\bf k}}
\newcommand{\vq}{{\bf q}}
\newcommand{\vp}{{\bf p}}
\newcommand{\vx}{{\bf x}}
\newcommand{\veci}{\vec{i}}
\title{ Coulomb gap in one-dimensional disordered electron systems}
\author{Hyun C. Lee}
\email{hyunlee@phys1.skku.ac.kr}
\affiliation{BK21 Physics Research Division and Institute of Basic Science,
 Department of Physics,\\
Sung Kyun Kwan University, Suwon, 440-746
Korea}
\date{\today}
\begin{abstract}
The density of states of one-dimensional disordered electron systems with long range Coulomb interaction is studied
in the weak pinning limit. The density of states is found to follow a power law with an exponent determined by
localization length, and this power law behavior is consistent with the existing numerical results.
\end{abstract}
\maketitle
\textit{Introduction}-Recently there have been much interests in the one-dimensional (1D) electron systems motivated by
the development of carbon nanotube technology.\cite{ebbesen}
In 1D electron systems electron-electron interactions play very important roles, 
leading to phases different from the conventional Fermi liquids.\cite{voit} 
Repulsive short range electron-electron interactions cause Luttinger liquids (LL),\cite{voit}
while long range Coulomb interaction (LRCI) is believed to
cause a Wigner crystal.\cite{schulz}
The Wigner crystal phase of spinful electrons is characterized by  quasi long range order of  $4 k_F $ charge density 
components.
\be
\label{wigner}
\langle \rho_{4 k_F}(x) \rho_{4 k_F}(0) \rangle  \sim e^{-\sqrt{\ln x}}.
\ee
Eq. (\ref{wigner}) should be compared with the power law dependence $x^{-\alpha}$ of LL.
Eq. (\ref{wigner}) also indicates the suppressed quantum fluctuations of charge densities, and this feature  manifest in the 
density of states (DOS)  $D(\omega)$.\cite{schulz} 
\be
\label{doswigner}
D(\omega) \sim \exp \Big[- {\rm const.}\,  \left( \ln  \frac{E_c}{|\omega|} \right )^{3/2} \Big],
\ee
where $\omega$ is measured from Fermi energy, and $E_c$ is a cutoff energy. Note that DOS of Eq. (\ref{doswigner}) decays faster
than any other power law.
For LL, DOS follows a power law $|\omega|^\gamma$ with a nonuniversal positive $\gamma$.
The physics of 1D Wigner crystal is analogous to that of charge density wave (CDW)\cite{cdw1,cdw2,gruner}
apart from  the presence of quantum fluctuations and LRCI.

Impurities, either a few or many,  change  the physical properties of 1D electron systems qualitatively.
For a single impurity in an electron system with repulsive interaction, the back scattering of electrons with the 
impurity  becomes strong at low energy, and it   effectively divides the system into two pieces.\cite{kane,akira}

For noninteracting electrons in random disorder (many impurities), all the states are known to be localized 
due to repeated back scattering.\cite{palee} DOS  for a disorder with a Gaussian distribution (with zero mean)
 can be calculated exactly.\cite{halperin,note}
\ba
\label{freecase}
D(E)&=& \frac{1}{\pi \sqrt{2E}},\;\;{\rm for}\;\; E \to \infty  \cr
D(E)&=& \frac{8(-E)}{3\pi}\,e^{-\frac{1}{12}(-8 E)^{3/2}},\;\;{\rm for } \;\;E \to -\infty.
\ea
In the strongly localized electron systems with LRCI, where the overlap of wave functions can be neglected, electrons can be treated 
classically and  DOS exhibits a Coulomb gap of the form:\cite{coulombgap}
\be
\label{vojta}
D(\omega) \sim \left( \ln  \frac{E_c}{|\omega|} \right )^{-1}.
\ee
We note that Eq. (\ref{vojta}) has been derived under the assumption that the localization length is much smaller than
the interparticle distance.
Both disorder and LRCI push  a 1D electron system to  classical limit but in a  different manner as is reflected in 
the form of DOS Eq. (\ref{doswigner}) and Eq. (\ref{vojta}). 
In this paper we report a result on the DOS at low energy for the 1D disordered electron system with LRCI
 when the localization length is larger than
the interparticle distance or interimpurity distance.[more precisely the weak pinning limit, see below ]
Following the analyses  on a pinned Wigner crystal by Maurey and Giamarchi \cite{giamarchi,gia2},  employing 
a simplified model, and using semiclassical approximation  we find that DOS follows a {\it power law} at low energy.
\be
\label{dos}
D(\omega) \sim |\omega|^{ \sqrt{1+\eta}/2},
\ee
where the exponent $\eta$  is basically determined by the localization length.[see Eq. (\ref{eta})]
This power law behavior is consistent with the existing numerical results.\cite{gun}
Eq. (\ref{dos}) is the main result of this paper.

\textit{Model}-
We consider a spinless electron system for simplicity. Such a system can be realized in organic chains and quasi-1D quantum
wires in strong magnetic field.\cite{material}
The Hamiltonian consists of three parts.
\ba
\label{H1}
H&=&H_0+H_{{\rm coul}}+H_{\textrm{imp}},\nonumber \\
H_0&=&v_F \int dx \Big[-i  \psi^\dag_{R }\partial_x \psi_{R }+i \psi^\dag_{L } \partial_x \psi_{L } \Big], \\
H_{{\rm coul}}&=& \int dx dy  \frac{V(x-y)}{2} \rho(x) \rho(y).
\ea
The operator $\psi_{R } (\psi_{L })$ is the right-moving (left-moving)  electron operator.
The continuum chiral electrons and lattice electron operators are related by
\be
c(x)=\sqrt{a}\Big[ e^{ i k_F x} \psi_R(x) +e^{- i k_F x} \psi_L(x) \Big],
\ee
where $a$ is lattice constant. 

$\rho_{R }=:\psi^\dag_{R } \psi_{R }:$ is the normal ordered right-moving
edge electron density operator ($\rho_{L }$ is similarly defined), and $\rho(x)=\rho_R(x)+\rho_L(x)$.
$V(x)=\frac{e^2}{\epsilon}\frac{1}{
\sqrt{x^2+d^2}}$ is the Coulomb interaction.  $d$ is the transverse size of quantum wire which we take to be the same  $a$
for simplicity.  $\epsilon$ is a dielectric constant.
The Coulomb matrix element is $V(k)=\frac{ 2 e^2}{\epsilon} K_0( a |k|) \sim \frac{2 e^2}{\epsilon} \ln \frac{1}{|k| a}$
for $|k|a > 1$ and $K_0$ is the modified Bessel function.
The impurity Hamiltonian is given by
\ba
\label{imp}
& &H_{\textrm{imp}}=\sum_x  W_I(x)  c^\dag(x) c(x) \nonumber \\
&=&\int dx W_I (x) \Big[  \rho(x)  +  e^{2 i k_F x}  \psi^\dag_L(x) \psi_R(x)+\textrm{H.c}\Big],
\ea
where $W_I(x)$ is the impurity potential.
The first term in the bracket of Eq. (\ref{imp}) is the forward scattering term which can be neglected compared to
back scattering at low energy.
The impurity potential $W_I(x)$ is chosen to be
\be
W_I(x)= \sum_{j} V_0  \delta(x-X_j),
\ee
where $X_j$'s are the random locations of impurities. 
The interacting electron systems can be bosonized in a standard way.\cite{voit,convention}
The phase fields  and bosonization formulas are given by
\ba
\label{formula}
\rho_R+\rho_L&=&\frac{1}{\pi}\partial_x \theta, \quad
\rho_R-\rho_L=\frac{1}{\pi}\partial_x \phi, \nonumber \\
\psi_{R}&=&\frac{e^{i \theta + i \phi}}{\sqrt{2\pi a}},\quad \psi_L=\frac{e^{-i \theta +i \phi}}{\sqrt{2\pi a}}.
\ea
The bosonized Hamiltonian of the system is
\ba
\label{bosonized}
H&=&\int dx  
\frac{v_F}{2\pi} \Big[ (\partial_x \theta)^2 + (\partial_x \phi)^2 \Big]  \nonumber \\
&+& \frac{1}{2\pi^2} \int dx dy \Big[ V(x-y)\, \partial_x \theta(x)  \, \partial_y \theta(y) \Big] \nonumber \\
&+& \sum_j V_0 \rho_0 \cos[2 k_F X_j + 2 \theta(X_j) ],
\ea
where $\rho_0$ is the average density of electrons.\cite{giamarchi}

\textit{Pinning Length}-The pinning length $L_0$ is a length scale over which the phase field $\theta(x)$ in the ground
state varies with $\delta \theta \sim \pi$ in order to take advantage of the impurity potential.\cite{giamarchi,gia2}
The pinning length corresponds to the localization length of the electron system.
The pinning length can be obtained by maximizing the energy gains from impurity, elastic, and Coulomb energies.
\cite{cdw2,giamarchi,gia2} Including the effect of quantum fluctuations using the self-consistent harmonic approximation,
the pinning length is given by
\be
\label{pinning}
L_0=\left(\frac{8e^2}{ \pi^2 V_0 \rho_0 \gamma n_i^{\frac{1}{2}}} \right)^{\frac{2}{3}}
\ln^{\frac{2}{3}}\left[\frac{1}{a} \left(\frac{8e^2}{ \pi^2 V_0 \rho_0 \gamma n_i^{\frac{1}{2}}} \right)^{\frac{2}{3}}
\right],
\ee
where $\gamma$ is a numerical constant characterizing short-range interaction and quantum fluctuations and $n_i$ is 
impurity concentration. The logarithmic factor is due to LRCI, which enhances the pinning length.
The enhanced pinning length implies the more rigid system and the more difficult pinning by impurties.
The system becomes much more ordered and the fluctuations around the ground state is much less important.
The expression for pinning length Eq. (\ref{pinning}) has been derived under the assumption of {\it weak pinning}
$L_0 \gg n_i^{-1}$. 

Beyond the pinning length scale the phase coherence of $\theta$ is lost, thus the correlation function of CDW operator 
is expected to decay exponentially\cite{gruner} ($|x-x^\prime| > L_0$)
\be
\label{correlation}
\langle \cos[2 \theta(x,\tau) ] \,\cos[2 \theta(x^\prime,\tau) ] \rangle \sim e^{-|x-x^\prime|/L_0}.
\ee
Eq. (\ref{correlation}) implies that the system breaks into segments whose typical length is given by the pinning length
$L_0$.
Thus at low energy we can consider a typical segment with length $L_0$  and calculate  DOS averaged over the segment.
Since the tunneling between segments is strongly suppressed at low energy we have to  fix the 
 value of $\theta$ at the boundary of a particular segment.\cite{kane}(Dirichlet boundary condition)

\textit{Approximate Low Energy Model}- Let us construct a model for a segment which is valid at low energy.
First of all the logarithmic divergence of Coulomb matrix element $V(k)$ is cut by $k \sim 1/L_0$. 
Then the Coulomb energy term can be expressed as 
\ba
\label{eta}
H_{{\rm coul}} &\sim &\eta\,\frac{v_F}{2\pi}\,\int_0^{L_0}\,
dx [ \partial_x \theta(x) ]^2, \nonumber \\
\eta &=&\frac{2 e^2 }{v_F \pi \epsilon} \,\ln \frac{L_0}{a}.
\ea
 The effective Hamiltonian of a segment becomes
\ba
\label{hamil}
H_{{\rm seg}}&=&\int_0^{L_0} dx \frac{v_F}{2\pi} \Big[
 (\partial_x \theta)^2 (1+\eta)+ (\partial_x \phi)^2  \Big] \nonumber \\
&+& V(\theta,X_j),
\ea
where $V(\theta,0< X_j < L_0)$ is some potential energy term 
which is optimized by a certain random {\it average} phase value $\theta_{{\rm opt}}$ and it depends on $\theta$ only through
$e^{ \pm 2 i \theta}$.[see the last line of Eq. (\ref{bosonized})]
Note that $\theta \to \theta+\pi$ is then a symmetry of the system.
 In the weak pinning limit we are considering the phase $\theta$ varies rather smoothly  in the range 
$|\theta-\theta_{{\rm opt}} | \leq \pi$ at low energy over a length scale $L_0$.

\textit{Calculation of DOS}-
The electron Green function in imaginary time is defined by 
\be
G_{R/L}(x,y,\tau_1-\tau_2,W_I)=-\langle T_\tau\,\psi_{R/L}(x,\tau_1) \psi^\dag_{R/L}(y,\tau_2) \rangle 
\ee
for a particular realization of impurities $W_I$.
After averaging ove impurities 
\be
\bar{G}_{R/L}(x-y,\tau_1-\tau_2)=\langle G_{R/L}(x,y,\tau_1-\tau_2,W_I) \rangle_{W_I}.
\ee
DOS is defined by
\be
D(\omega)=-\frac{1}{\pi}\,\textrm{Im}\Big[ \int_0^\beta d \tau e^{i \omega \tau} \bar{G}_{R/L}(0,\tau) \Big]_{
i\omega \to \omega+ i \delta}.
\ee
The above expressions are formally exact but difficult to calculate. 
Here instead we will calculate the electron Green function at a point $y$ in the segment (away from the boundary) using the Hamiltonian
Eq. (\ref{hamil}).
\be
\langle G_{R/L}(y,y,\tau_1-\tau_2 ) \rangle_{W_I}.
\ee
For the computation of electron Green function, the Lagrangian formulation is more convenient.
\be
S_{{\rm seg}}=\int_0^{L_0} dx \int_0^\beta\,d \tau \frac{i}{\pi} \partial_\tau \theta\,\partial_x \phi 
+\int_0^\beta d\tau H_{{\rm seg}}.
\ee

Using the bosonization formula   $G_{R/L}(y,\tau_1-\tau_2 )$ can be expressed as 
\ba
\label{green}
& &G_R(y, \tau_1- \tau_2)=\frac{\int D[\theta,\phi]\,e^{-S_{{\rm seg}}-S_{\textrm{ex}}}}{\int D[\theta,\phi]e^{-S_{{\rm seg}}}}, \nonumber \\
& &S_{\textrm{ex}}= i \int dx d\tau \Big[ \theta(x,\tau) + \phi(x,\tau) \Big] J(x,\tau), \nonumber \\
& & J(x,\tau)= \delta(x-y) \Big [ \delta(\tau-\tau_1)- \delta(\tau-\tau_2)  \Big ].  
\ea
In Eq. (\ref{green}) the dual phase field $\phi$ can be integrated out explicitly. [Remember that the potential term $V(\theta,X_j)$
depends only on $\theta$ not on $\phi$.] 
The electron Green function becomes
\be
G_R(y, \tau_1-\tau_2)=\frac{\int D[\theta]\,e^{-S_\theta-S_{J \theta}-S_J}}{\int D[\theta]\,e^{-S_\theta}}.
\ee
\ba
S_\theta&=&\frac{u}{2\pi K}\,\int_0^{L_0} dx \int_0^\beta  d\tau\,\Big[
\frac{1}{u^2}(\partial_\tau \theta)^2+(\partial_x \theta)^2 \Big] \nonumber \\
&+& \int_0^\beta d \tau \, V(\theta,X_j), \nonumber \\
K&=&\frac{1}{\sqrt{1+\eta}},\quad u=v_F \sqrt{1+\eta}.
\ea
\ba
\label{source}
S_{J\theta }&=&-\frac{1}{2 K}\,\sum_{\omega,k} \frac{\omega}{u k}
\Big[ J(-i \omega, -k) \theta(i \omega, k)+{\rm H.c.} \Big]  \nonumber \\
&+&i \int d x d \tau \,\theta(x,\tau) \,J(x,\tau).
\ea
\be
\label{jjj}
S_J=\frac{\pi}{2 K} \sum_{\omega,k}\, \frac{J(-i\omega,-k) J(i\omega,k)}{ u k^2}.
\ee
Because our system is in the deep classical regime the dominant low energy processes would come from the quantum tunneling
between classical vacua. Remembering that the phase field $\theta$ is an angular variable, the classical vacua are charerized by
$\theta_{{\rm vac}} \sim \{\theta_{{\rm opt}}+ \pi n, n={\rm integer} \}$.
This is because the potential $V(\theta,X_j) $ depends on $\theta$ only through the form $e^{\pm 2i \theta}$,  so that
it possesses the  symmetry
\be
\label{sym}
\theta \to \theta+\pi
\ee 
Since the phase field $\theta$ varies on the order $\pi$ in each segment one can expect that the most domiant process would be
the quantum tunneling between  $\theta_{{\rm opt}}$ and $\theta_{{\rm opt}} \pm \pi$ vacua.
The quantum tunneling processes between classical vacua can be described by the solutions of classical equations of motion
in {\it imaginary time}, which are also called {\it instantons}.\cite{polyakov}

Now we have to find the classical {\it (imaginary) time dependent} solution which minimizes $S_\theta+S_{J \theta}$.
Since $S_{J \theta}$ is linear in $\theta$ it plays a role of the external source field. 
We argue below that the first term of  $S_{J \theta}$ [ Eq. (\ref{source}) ] is 
the source for the vortex and anti-vortex configuration of the (angular) phase field $\theta$.
For the moment let us neglect the potential term $V(\theta,X_j)$. Then the phase field $\theta$ and its dual phase field $\phi$
 are related by
the Cauchy-Riemann equation\cite{tsvelik} [we set $u=1$ for simplicity from now on]
\be
\label{cauchy}
\partial_\mu \phi=-i  \epsilon_{\mu \nu} \partial_\nu \theta,\;\;\mu,\nu=\tau, x,
\ee
where $\epsilon_{\mu \nu}$ is  a totally antisymmetric tensor. The first term  Eq. (\ref{source}) stems from  
the source of dual field  $\phi$.[see Eq. (\ref{green})] Including  only one source at $(\tau_1,y)$
the $\phi$ field satisfies the classical equation of motion.
\be
\label{eom}
-(\partial_\tau^2+  \partial_x^2) \phi(x,\tau)=i \pi \delta(x-y) \delta(\tau-\tau_1).
\ee
Combining Eq. (\ref{cauchy}) and Eq. (\ref{eom})  we deduce
\be
\label{singular}
 \pi \delta(x-y) \delta(\tau-\tau_1)= \epsilon_{\mu \nu}\partial_\mu \partial_\nu \theta(x,\tau),
\ee
Naively the right hand side of Eq. (\ref{singular}) vanishes but this is not necessarily true for the topologically
singular configuration such as vortex.
The vortex nature of $\theta$ is demonstrated by the calculation of vorticity using Eq. (\ref{singular}) and Stokes theorem.
\be
\label{vorticity}
\oint d x_\mu\,\partial_\mu \theta=\int d^2 x \epsilon_{\mu \nu}\,\partial_{\mu} \partial_{\nu} \theta=\frac{1}{2} 2\pi,
\ee
which characterizes the vortex nature of $\theta$.\cite{zee} The unusual factor of 1/2 is due to the aforementioned
symmetry $\theta \to \theta +\pi$.
Remembering that we have another souce at $(\tau_2,y)$ with a charge opposite to that at $(\tau_1,y)$[see the last line
of Eq. (\ref{green})], we conclude that 
the vortex-antivortex configuration of $\theta$ phase field  will dominate the electron Green function at long time 
in the classical regime. Before proceeding to the explicit vortex-antivortex instanton solution, let us discuss another
classical field configurations generated by the  second source term of Eq. (\ref{source}).
It satisfies the standard Laplace equation in two dimension with delta function sources, and its explicit form is given by 
\be
\label{second}
\theta_0(\tau,x) \sim \frac{  K }{4 }  \ln \frac{(x-y)^2+(\tau-\tau_1)^2}{(x-y)^2+(\tau-\tau_2)^2}.
\ee
In the configuration of Eq. (\ref{second}) the range of variation of phase field $\theta$ is much large than $\pi$
for long time limit $|\tau_1-\tau_2| \to \infty$. Thus the configuration $\theta_0(x)$ is suppressed by 
the potential term $V(\theta,X_j)$, and  it also does not exhibit the angular nature of phase fields.
This is an analogue of the spin-wave degrees of freedom of XY model.\cite{polyakov}  Thus,
the configuration of Eq. (\ref{second}) does not contribute the electron Green function in long time limit.

The explicit  vortex-antivortex  solution  in complex coordinate is given by \cite{polyakov}
\be
\label{vortexsol}
e^{i 2 \theta(z)}=\frac{z-( \tau_1 + i y)}{z-(\tau_2+i y)},\;\; z=\tau+i x
\ee
Substituting Eq. (\ref{vortexsol}) into the  action $S_\theta$ we obtain
\ba
\label{result}
G_R(y,\tau_1-\tau_2)& \sim& \exp[-\frac{1}{  2 K} \ln \frac{|\tau_1-\tau_2|}{\tau_0} \Big ], \nonumber \\
&=&\Big [\frac{|\tau_1-\tau_2|}{\tau_0} \Big ]^{-1/2K},
\ea
where $\tau_0$ is the short time cut-off provided by the contributions from the vortex cores.
Eq. (\ref{result}) implies Eq. (\ref{dos}).

To complete our analysis let us compute $S_J$  Eq (\ref{jjj}). 
The  divergent $k$ summation at small $k$ should be cut by $1/L_0$. The frequency summation diverges at
large frequency (short time) which is cut by some high energy scale such as $v_F n_i$.\cite{akira} 
Thus, $S_J$ does not  contribute to the electron Green function in long time limit.

According to Eq. (\ref{dos}) the exponent of  DOS  is larger for the longer localization length,
 which is consistent with 
the numerical results by Jeon {\it et al.}.\cite{gun}  
The exponent $\eta$ can be expressed as
\be
\eta = \frac{1}{137}\times \frac{2 c}{v_F} \times \frac{1}{\epsilon} \times \ln \frac{L_0}{a},
\ee
where $c$ is the speed of light. In the weak pinning limit it is reasonable to take  $ \ln \frac{L_0}{a}$ to be 4-6.
Choosing a typical Fermi velocity near $10^7 {\rm cm}/{\rm sec}$ and $\epsilon \sim 1-5$ for a quantum wire,
 we can estimate the exponent of DOS $\sqrt{1+\eta}/2$  to be around 3 - 6 which is also consistent
with the exponents obtained from the numerical studies.\cite{gun}

More precisely the Eq. (\ref{result}) should be averaged over the impurity configurations in the {\it segment}.
Inside the segment the phase field varies smoothly in space and the impurity average is not expected to bring in any singular effects
which would invalidate Eq. (\ref{result}), since no infrared divergence can occur through impurity averages for a finite
segment.

Sufficiently strong short range interaction can also push the system to classical regime $(K << 1)$, where 
the result Eq. (\ref{dos}) is applicable. On the contrary, the result Eq. (\ref{dos}) 
is not applicable to the case of noninteracting disordered electrons ($ K=1 $) 
since the condition of the classical limit $(K << 1)$ is not met.
Indeed the DOS of noninteracting disordered electrons is finite at the Fermi level (See Eq. (\ref{freecase}))
contrary to Eq. (\ref{dos}).

\textit{Summary}-
We have studied  analytically the DOS of the disordered 1D electron system interacting via long range Coulomb interaction
employing a simplified model and semiclassical approximation. The DOS is found to follow a power law with a nonuniversal exponent 
which is basically determined by localization length. The power law dependence is also consistent with 
the existing numerical result.

\begin{acknowledgments}
We are grateful to G. S. Jeon for useful comments.
This work was  supported by the Korea Science and Engineering
Foundation (KOSEF) through the grant No. 1999-2-11400-005-5, and by the 
Ministry of Education through Brain Korea 21 SNU-SKKU Program.
\end{acknowledgments}


\begin{thebibliography}{}
\bibitem{note}  $E=\big[\frac{\hbar}{D m^{1/2}} \big ]^{-2/3} \omega$, where $m$ is a mass of electron and $D$ is defined by
the distribution of random potential $\langle V(x) V(y) \rangle = D \delta(x-y)$.
\end{thebibliography}
\end{document}